# A metallic mosaic phase and the origin of Mott insulating state in 1T-TaS$_2$

Liguo Ma[1,3], Cun Ye[1,3], Yijun Yu[1,3], Xiu Fang Lu[2,3], Xiaohai Niu[1,3], Sejoong Kim[4]
Donglai Feng[1,3], David Tománek[5], Young-Woo Son[4], Xian Hui Chen[2,3], Yuanbo Zhang[1,3*]

[1]*State Key Laboratory of Surface Physics and Department of Physics, Fudan University, Shanghai 200433, China*

[2]*Hefei National Laboratory for Physical Science at Microscale and Department of Physics, University of Science and Technology of China, Hefei, Anhui 230026, China*

[3]*Collaborative Innovation Center of Advanced Microstructures, Nanjing 210093, China*

[4]*Korea Institute for Advanced Study, Hoegiro 87, Dongdaemun-gu, Seoul, Korea*

[5]*Physics and Astronomy Department, Michigan State University, East Lansing, Michigan 48824, USA*

*Email: zhyb@fudan.edu.cn

**Electron-electron and electron-phonon interactions are two major driving forces that stabilize various charge-ordered phases of matter. The intricate interplay between the two give rises to a peculiar charge density wave (CDW) state[1–3], which is also known as a Mott insulator[2], as the ground state of layered compound 1T-TaS$_2$. The delicate balance also makes it possible to use external perturbations to create and manipulate novel phases in this material[4,5]. Here, we study a mosaic CDW phase induced by voltage pulses from the tip of a scanning tunneling microscope (STM)[6], and find that the new phase exhibit electronic structures that are entirely different from the Mott ground state of 1T-TaS$_2$ at low temperatures. The mosaic phase consists of nanometer-sized domains characterized by well-defined phase shifts of the CDW order parameter in the topmost layer, and by altered stacking relative to the layer underneath. We discover that the nature of the new phases is dictated by the stacking order, and our results shed fresh light on the origin of the Mott phase in this layered compound.**

When the correlation between electrons become predominant, the interaction may lead to the localization of electrons in materials with half-filled energy bands, and turn the otherwise metallic systems into insulators[7]. Such insulator (the so called Mott insulator, MI), therefore, serves as an ideal starting point for the study of strongly correlated electron systems. Indeed, MI and its transition to metallic state (commonly referred to as metal-insulator transition) form the basis of our understanding of various magnetic phenomena[8,9] and high-temperature superconductivity[9–11].

The correlation effects play an important role in layered transition metal dichalcogenide 1T-TaS$_2$, which turns into a MI after a series of CDW phase transitions as the temperature is lowered [1–3,12]. The Mott-insulating ground state of 1T-TaS$_2$, however, differs from typical MIs in that it resides inside a commensurate CDW (CCDW) state. As a result the localization centers in 1T-TaS$_2$ are CDW superlattices, instead of atomic sites found in conventional MIs; there is also no apparent magnetic ordering accompanying the MI formation in 1T-TaS$_2$ (ref. 12–15). Meanwhile, because of the close proximity of the various competing charge ordered phases in energy, external perturbations can effectively modulate the CCDW (and thus the Mott phase) in 1T-TaS$_2$ and induce a myriad of phase transitions[4–6,16–27]. 1T-TaS$_2$ is therefore well suited to be a testbed for MI and other related strongly correlated phases.

In this study, we use voltage pulses from an STM tip to create a mosaic CDW state out of the Mott ground state of 1T-TaS$_2$ following a procedure described in ref. 6, 28. We found that the mosaic state exhibits metallic behavior that is fundamentally different from the parent MI state. Atomically resolved mapping of the mosaic metallic (MM) phase uncovers the root of such difference: each domain in the top layer of the mosaic phase is characterized by well-defined phase shift of the CDW order parameter with respect to neighbouring domains, and to the layer underneath; the altered stacking of CDW superlattice dictates whether the new phase is MI or MM phase. Our results therefore provide fresh insight to the origin of the Mott phase in 1T-TaS$_2$ that has so far been shrouded in controversies[29–31]. Moreover, we find that the MM phase created at low temperature is metastable in nature: it switches back into the Mott phase after a thermal cycle. Such observation links the MM phase to the metastable phases of 1T-TaS$_2$ induced by ultra-fast laser pulses[22] and current excitation[24–27]. Our result may therefore provide a microscopic understanding for those novel phases that are of importance in practical applications.

1T-TaS$_2$ bulk crystal has a layered structure, with each unit layer composed of a triangular lattice of Ta atoms, sandwiched by S atoms in an octahedral coordination. The Ta lattice is susceptible to in-plane 'David-star' deformation where 12 Ta atoms contract toward a central Ta site[1] as illustrated in Fig. 1a. Below 180 K the crystal enters a CCDW phase, where the David-stars become fully interlocked, forming a $\sqrt{13} \times \sqrt{13}$ superlattice[12]. Such commensurate lattice modulation is accompanied by electronic reconstructions, which split the Ta 5d-band into several submanifolds, leaving exactly one conduction electron per David-star. Strong electron-electron interaction further localize these electrons, and leads to a MI ground state[12].

Displayed in Fig. 1b is the surface of 1T-TaS$_2$ in the MI-CCDW phase imaged by STM at 6.5K. A prominent CCDW superlattice is clearly resolved with each of the bright spots corresponding to a CDW cluster. Close examination of individual cluster reveals the position of the S atoms in the topmost layer (Fig. 1b inset), which bulges vertically to accommodate the distortion of the Ta lattice, and forms truncated triangles located right above the David-stars (marked red, Fig. 1b inset). The in-plane CDW charge modulation $\delta\rho$ can be described by a set of complex order parameters, $\phi_i(\boldsymbol{r}) = \Delta_i(\boldsymbol{r}) \cdot e^{\theta_i(\boldsymbol{r})}$, such that $\delta\rho = Re[\sum_{i=1,2,3} \exp(i\boldsymbol{Q}_i\boldsymbol{r}) \cdot \phi_i(\boldsymbol{r})]$. Here $\boldsymbol{Q}_i$ ($i = 1,2,3$) are the three in-plane reciprocal lattice vectors associated with the $\sqrt{13} \times \sqrt{13}$ supermodulation, and $\theta_i(\boldsymbol{r})$ and $\Delta_i(\boldsymbol{r})$ represent the phase and amplitude of the CDW charge order. The regular triangular superlattice seen in Fig. 1b indicates a ground state with uniform $\phi_i$ in 1T-TaS$_2$.

The MI ground state makes drastic transition to a mosaic state when subjected to a voltage pulse applied across the tip-sample junction at low temperatures. A patch of such pulse-generate mosaic state is presented in Fig. 1c. Inside the patch the originally homogeneous CCDW superlattice disintegrates into nanometer-sized domains separated by well-defined domain wall textures. Such mosaic phase is distinctly different from the nearly commensurate CDW (NCCDW) phase existing at higher temperatures (Ref. 3,32 and Supplementary Section VI), but shows strong similarity to the super-cooled NCCDW state at low temperatures (Supplementary Section VII). Within the domains the commensurate David-star configuration is strictly preserved (Fig. 1c inset), whereas the phase of the CDW order $\theta_i(\mathbf{r})$ undergoes abrupt change across the domain walls, which we shall discuss later.

The electronic structure of the new mosaic state is fundamentally different from that of the original MI ground state. Whereas the pristine CCDW ground state is a MI featuring a 430 meV Mott-Hubbard gap (Fig. 2c, black), the mosaic state is of a metallic nature with finite local density of states (LDOS) around the Fermi level both inside the domains (Fig. 2c, red) and on the domain walls (Fig. 2c, blue). Such a distinction is clearly captured by the $dI/dV$ spectral waterfall acquired across a MM-MI interface shown in Fig. 2b, where a sharp metal-insulator transition occurs within a superlattice unit cell. The transition from MM to MI is accompanied by prominent deformation of the Hubbard bands on the MI side, which we attribute to the combined space charging and tip-induced band bending effect similar to that at a semiconductor-metal interface (Supplementary Section V).

Even though the MM state appears stable at low temperatures, we find that the state is in fact metastable in nature. Fig. 3a displays a typical pulse-induced MM patch surrounded by the pristine MI state. No change of the MM patch was observed after weeks of intensive imaging and spectroscopic measurements at $T = 6.5$ K. Upon increasing the temperature, however, the MM patch becomes unstable and the domain structure melts away. Fig. 3b display the same area of the sample surface as shown in Fig. 3a, but at an elevated temperature of $T = 46$ K. The dense aggregation of domain walls dissolves, leaving an ordered CCDW superlattice decorated with sparse boundary lines, part of which traces the low-temperature domain walls (Fig. 3b, dashed lines). Meanwhile, the MI insulating behavior fully recovers over the entire surface except on those boundary lines. Cooling down the sample again does not bring back the MM state, and the MI state (with boundary lines) persists to low temperatures. Such hysteretic bebaviour unambiguously demonstrates that the MM state is metastable. The metastable nature of the MM state is further corroborated by the state's fragility at elevated temperatures: the border of the metallic state gradually recedes when perturbed by repeated scanning of an STM tip at $T = 30$ K (tunneling voltage and current is $V_t = 150$ mV and $I_t = 10$ pA, respectively) (Fig. 3c). Finally, we note that the MM phase observed in our experiment may be intimately linked to the metastable metallic states induced by various macroscopic techniques, including both optical excitation[22] and carrier injection[24–27], from the ground state of 1T-TaS$_2$. Our STM study may therefore provide crucial microscopic understanding of those phases for the first time.

A key feature of the MM state is the constant phase of the order parameter $\theta_i$ inside each domain, and abrupt change in $\theta_i$ across the domain walls. Shown in Fig. 4a and b are two most common types of domain wall observed in the MM state. The atomically resolved STM images recorded at low bias ($V_t = 15$ mV) enable us to precisely determine the positions of the David-stars on the domain wall. It turns out that the superlattices of David-stars in the neighbouring domains are shifted against each other by a lattice vector of the underlying two-dimensional (2D) atomic crystal, $\boldsymbol{T}$, so that the phase difference between the two domains can be written as

$$\Delta\theta_i = \boldsymbol{Q}_i \cdot \boldsymbol{T} \qquad (1)$$

$\boldsymbol{T}$ takes the value of $-\boldsymbol{a} + \boldsymbol{b}$ and $-\boldsymbol{b}$ ($\boldsymbol{a}$ and $\boldsymbol{b}$ are basis vectors of 1T-TaS$_2$ 2D crystal) for domain walls shown in Fig. 4a and 4b, respectively. As a result two columns of David-stars form the domain wall in an edge-sharing (Fig. 4a) or corner-sharing (Fig. 4b) configuration. No rotation of the David-star triangular lattice was observed on the domain wall, nor was any defect in the underlying atomic crystal.

In fact, surveys of the MM state reveal that equation (1) describes all the domain walls observed in our experiment (Supplementary Section III), which enables us to completely determine the phase configuration of the MM domains. The relative translation $\boldsymbol{T}$ in general takes the form $\boldsymbol{T} = m\boldsymbol{a} + n\boldsymbol{b}$, where $m$ and $n$ take integer values such that a $\boldsymbol{T}$ sitting at the central Ta atom runs through twelve other Ta on the same David-star (Fig. 4c, Supplementary Section III). There are therefore twelve possible types of domain walls with associated phase difference $(\Delta\theta_1, \Delta\theta_2, \Delta\theta_3) = 2\pi(3m + n, -4m + 3n, m - 4n)/13$. Equivalently, the twelve possible $\boldsymbol{T}$ can be labeled by the Ta atom that characterizes the translation (Fig. 4c). Here the Ta atoms (and therefore the $\boldsymbol{T}$) are numbered **0** ... **12** following the convention adopted in ref. [33], which has the added advantage that two consecutive translations are represented by the difference of the two numbers. Armed with above analysis, we are able to completely delineate the phase configuration of an MM state, such as the one shown in Fig. 4d. Here the phase of all the domains is referenced to the domain **0**, and the relative phase shift between two domains is readily unscrambled by taking the difference of their domain numbers. Finally, we note that out of the twelve possibilities only four types of domain walls are observed in our experiment with varying frequency of occurrence (Supplementary Section III), indicating subtle differences in energy associated with each type of domain walls.

The domain and associated phase structure distinguishes the MM state from its parent MI state, even though the two states share the same CCDW superlattice. There are two main points to notice. First, the MI state is parasitic to the CCDW[5], rather than a requisite as suggested by some of the previous works[34,35]. Second, phase shift of the CCDW order in one atomic layer implies a shifted CCDW superlattice relative to other layers. In fact, not only do we see phase shifts in the topmost layer, clear signature of random domain wall networks are also observed in the second layer (Fig. 5a). One example of such domain wall is shown in Fig 5c, where a flat monolayer H-$TaS_2$ patch (also induced by the same voltage pulse) enables us to see through the top layer and resolve the domain wall's atomic structure. We find that the domain wall in the second layer also corresponds to a phase shift described by equation (1) (Fig. 5d). The presence of randomly distributed domains and phases in two adjacent layers therefore incurs a randomized stacking of the CCDW superlattices.

We are now poised to address the central question: how does the MI state emerge from CCDW, against competing metallic state in 1T-$TaS_2$? An important clue comes from rare occurrences such as the one shown in Fig. 5b, where a small patch of MI state appears inside metallic random MM domains. Close examination reveals that the MI patch is surrounded partly by the topmost-layer domain walls, and partly by the domain wall in the second-layer. Such a domain wall configuration leads to a distinct stacking order of CCDW superlattice that restores the MI state inside the patch, in contrast to the surroundings domains still in the MM state. We therefore conclude that interlayer stacking plays a decisive role in determining the electronic structure of 1T-$TaS_2$ ground state.

The importance of the interlayer stacking can be understood from a three-dimensional Hubbard model with intra- and inter-layer hopping taken into account. The one-band Hamiltonian of 1T-TaS$_2$ CCDW ground state can be written as:

$$H = - \sum_{<ij>,\sigma} t_{ij}\left(c^\dagger_{i\sigma} c_{j\sigma} + H.c.\right) + U \sum_i n_{i\uparrow} n_{i\downarrow} \qquad (2)$$

where $t_{ij}$ is the effective hopping between the David-stars, and $U$ the on-site Coulomb repulsion on one David-star. Because of the flat 'pancake' shape of the David-star, the inter-layer distance of the CCDW superlattice (5.9 Å) is significantly shorter than the in-plane separation between the centers of neighbouring David-stars (12.1 Å; ref. 2). Various experimental and theoretical studies[30,31,36,37] have suggested the importance of interlayer coupling. Indeed, our ARPES measurements on the pristine crystal reveals a bandwidth of $W \sim 50$ meV for the lower Hubbard band in both $\boldsymbol{k}_z$ and $\boldsymbol{k}_\parallel$ direction (Supplementary Section I). This observation indicates that the effective out-of-plane hopping factor $t_\perp$ is comparable to its in-plane counterpart $t_\parallel$. We note that $U/W \sim 8$ for pristine 1T-TaS$_2$, a typical value for a Mott insulator ground state.

A MI to metal transition occurs upon increasing the bandwidth $W$ with respect to the Hubbard $U$, where $W$ is determined by the effective hopping factors $t_\perp$ and $t_\parallel$ (with coordination number taken into account). As the stacking order of the CCDW is varied, the variation in the separation between the David-stars in neighbouring layer (as well as coordination number) could bring drastic change in $t_\perp$. We speculate that in certain stacking configurations, the bandwidth $W$ is driven beyond certain critical value, and the MI insulator to metal transition becomes a possibility. Such a speculation is supported by recent DFT calculations suggesting that, with an altered interlayer stacking sequence, $t_\perp$ may experience an order of magnitude increase, and brings $W$ to the same order of magnitude as $U$ (ref. 31). The rare occurrence of the MI domain among the randomly stacked MM domains (Fig. 5b) implies that only a small number of stacking order yields MI states with the rest metallic in nature. The exact inter-layer stacking order of the various states, however, is not determined in our experiment. Here we note that even the stacking order in the pristine CCDW ground state remains elusive[33,38–41], and we call for further experimental and theoretical work to clarify the exact stacking order of the various charge-ordered states.

In summary, we studied a mosaic, metallic state induced from the Mott insulating CCDW phase in 1T-TaS$_2$ by voltage pulses. The mosaic phase features a fragmented in-plane phase distribution of the CDW order parameter, and exhibits metallic behaviour. We discovered that the relative phase shift between adjacent layers leads to an altered local stacking order, which dictates whether the resulting structure is a MI or a metal. Our results therefore shed fresh light to the origin of the MI state in the CCDW ground state of 1T-TaS$_2$, and uncover the interlayer coupling as the root. In addition, the MM phase bears strong similarities to the metastable metallic states induced by various external perturbations such as ultra-fast laser pulses[22] and current excitations[24–27]. Our study may provide a microscopic understanding for those novel phases that are of importance in practical applications.

*Note added*: During the preparation of our manuscript, we became aware of a related experiment on 1T-TaS$_2$ (ref. 42).

**Methods**

**Sample preparation and STM measurements.** High-quality 1T-TaS$_2$ single crystals were grown using a standard chemical vapor transport method. The samples were cleaved in high vacuum at room temperature, and subsequently cooled down for STM measurements. STM experiments were performed in a commercial low-temperature STM (Createc Fischer & Co. GmbH) operated in ultrahigh vacuum. Electrochemically etched polycrystalline tungsten tips were used for all our measurements. The STM topography was taken in the constant-current mode, and the $dI/dV$ spectra were collected using a standard lock-in technique with a modulation frequency of 789.1 Hz. Before we opened the feedback loop to apply voltage pulses, the tip was parked above the sample surface under typical tunneling condition of $V_t = 0.3 \text{ V} - 0.5 \text{ V}$ and $I_t = 0.3 \text{ nA} - 1 \text{ nA}$. The pulse duration was fixed at 50 ms.

**Acknowledgements**

We thank S.W. Cheong and Y.-H. Cho for providing some of the 1T-TaS$_2$ crystals used in this study, and P. Kim, J. Guan for discussions. L.M., C.Y., Y.Y. and Y.Z. acknowledge financial support from the National Basic Research Program of China (973 Program; grants nos. 2011CB921802 and 2013CB921902), and from the NSF of China (grant no. 11425415). X.F.L., Y.J.Y. and X.H.C. are supported by the 'Strategic Priority Research Program (B)' of the Chinese Academy of Sciences (grant no. XDB04040100) and the NSF of China (grant no. 11190021). Y-W.S. is supported by the NRF of Korea grant funded by MEST (no. 2015001948). D.T. acknowledges the hospitality of Fudan University while performing this research.

**Figure captions**

**Figure 1 | MM state induced from CCDW ground state in 1T-TaS$_2$ by voltage pulses. a,** Schematic of a monolayer 1T-TaS$_2$ crystal viewed from top. Only the topmost layer of S atoms are shown here for clarity. The interlocked clusters of Ta atoms (the 'David-star') in the CCDW state are sketched in purple, and the gray lines outline the S atoms accompanying each David-star. ***A*** and ***B*** are in-plane basis vectors of the CCDW superlattice, whereas ***a*** and ***b*** are the basis of the underlying atomic lattice. **b,** STM topography of the cleaved surface of pristine 1T-TaS$_2$. The $\sqrt{13} \times \sqrt{13}$ triangular CCDW superlattice is resolved at 6.5K. Inset: A zoomed-in view of the CCDW order. Individual S atoms are resolved on each cluster, which enable us to locate the Ta atoms underneath in David-star formation (red). **c,** Topographical image of an

MM patch generated by a 2.8 V voltage pulse in a MI background. Inset: CCDW charge order of the MM state with the same David-star formation. STM images were taken under the tunneling condition $V_t = 0.5$ V and $I_t = 0.1$ nA (main panels), or $V_t = 0.04$ V and $I_t = 2$ nA (insets).

**Figure 2 | Electronic structure of the MM state. a,** STM topography of the MM state (left) interfaced with MI state (right). Image recorded under tunneling condition $V_t = 0.3$ V and $I_t = 0.1$ nA. **b,** Differential conductance ($dI/dV$) as a function of sample bias (vertical axis) and distance (horizontal axis) measured along the yellow line in **a**. **c,** Spatially averaged $dI/dV$ spectra acquired in MI state (black), MM domains (red) and domain walls (blue). Curves are vertically shifted for clarity. The squares signify the spectral peaks corresponding to the submanifolds associated with the CCDW formation[37]. The triangles label the position of the Hubbard bands in MI state (black), or the edges of the V-shaped DOS suppression in MM state which are reminiscent of the Hubbard bands (red and blue).

**Figure 3 | The Metastable nature of the MM state. a,** Low-temperature STM topography of the MM state recorded at $T = 6.5$ K. **b,** Topographical image taken over the same area as in **a**, but at an elevated temperature of $T = 46$ K. The ramping rate of temperature was kept under 0.2 K/min. Part of the sparse residual boundaries (marked by red dash lines) are inherited from original domain walls in the low-temperature MM state as shown in **a**. **c,** An MM patch gradually converted to MI state by repeated scanning at $T = 30$ K. Panels from left to right are in chronological order, and the data were taken over a 6 min period. STM scanning condition: $V_t = 0.15$ V and $I_t = 0.01$ nA.

**Figure 4 | In-plane domain/phase configuration of the CCDW order in the MM state. a,** and **b,** STM topography two common types of CDW domain walls recorded at varying sample biases. The atomically resolved images at low bias enabled us to establish the structure model of both types of domain walls (bottom two panels). **c,** David-star of Ta atoms in the unit cell of the CCDW superlattice. The atoms are numbered following the convention described in ref. 33. Each number represents one of the twelve possible relative translations (and thus phase differences) of the CCDW superlattices in neighbouring domains. Two consecutive translations can be represented by the difference of the two numbers. **d,** Topography of an MM state showing multiple domains. The phase of each mosaic domains (relative to the domain '**0**') are determined from analysis of domain walls using procedures similar to **a**. The phases are coded by the numbers defined in **c** (see text).

**Figure 5 | Altered interlayer stacking order of the CCDW superlattice in the MM state. a,** STM topography of a 50 nm × 50 nm MM area. STM imaging condition: $V_t = 0.2$ V and $I_t = 0.1$ nA. Apart from the well-defined domain walls in the top layer, domain walls in the layer underneath are clearly resolved as networks of random filamentary features. Here an inverted color scale is adopted for better contrast. **b,** An MI patch (bright region) surrounded by MM domains in the MM state. The patch is encircled by domain walls in either the top layer (left half) or the second layer (right half). STM imaging condition: $V_t = 0.3$ V and $I_t = 0.02$ nA. **c,** A triangular patch of H-phase $TaS_2$ accidently created in 1T-$TaS_2$ by a voltage pulse. The H-phase is induced only on the topmost atomic layer, and the flat surface enabled us to see through the top layer and to observe the $\sqrt{13} \times \sqrt{13}$ CCDW superlattice as well as domain walls in

the layer underneath[6,16]. **d,** Zoomed-in STM image of the area marked in **c** showing atomically resolved domain wall structure in the second layer. The domain wall configuration corresponds to the type **2** discussed in Fig. 4c.

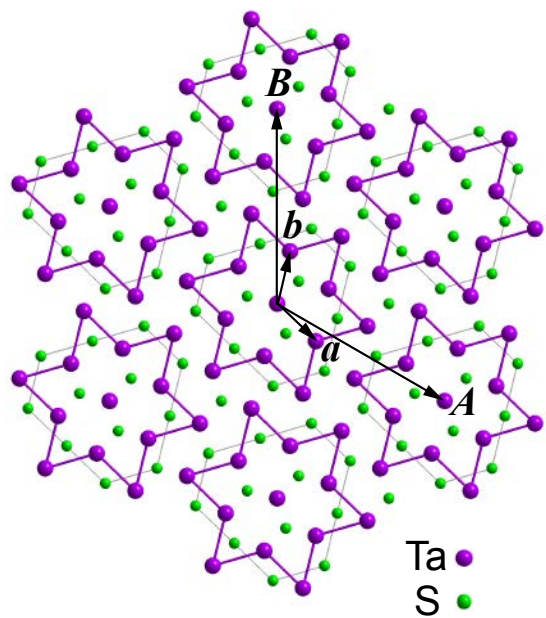 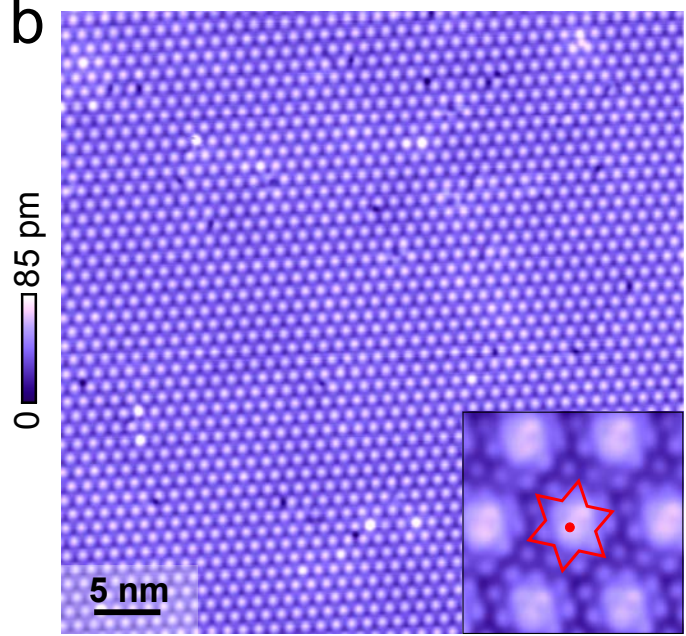 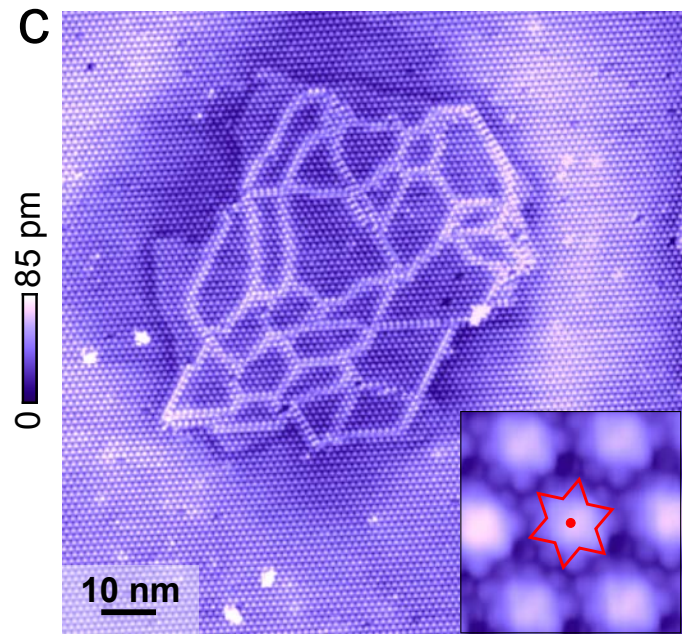

Liguo Ma *et al*. Figure 1

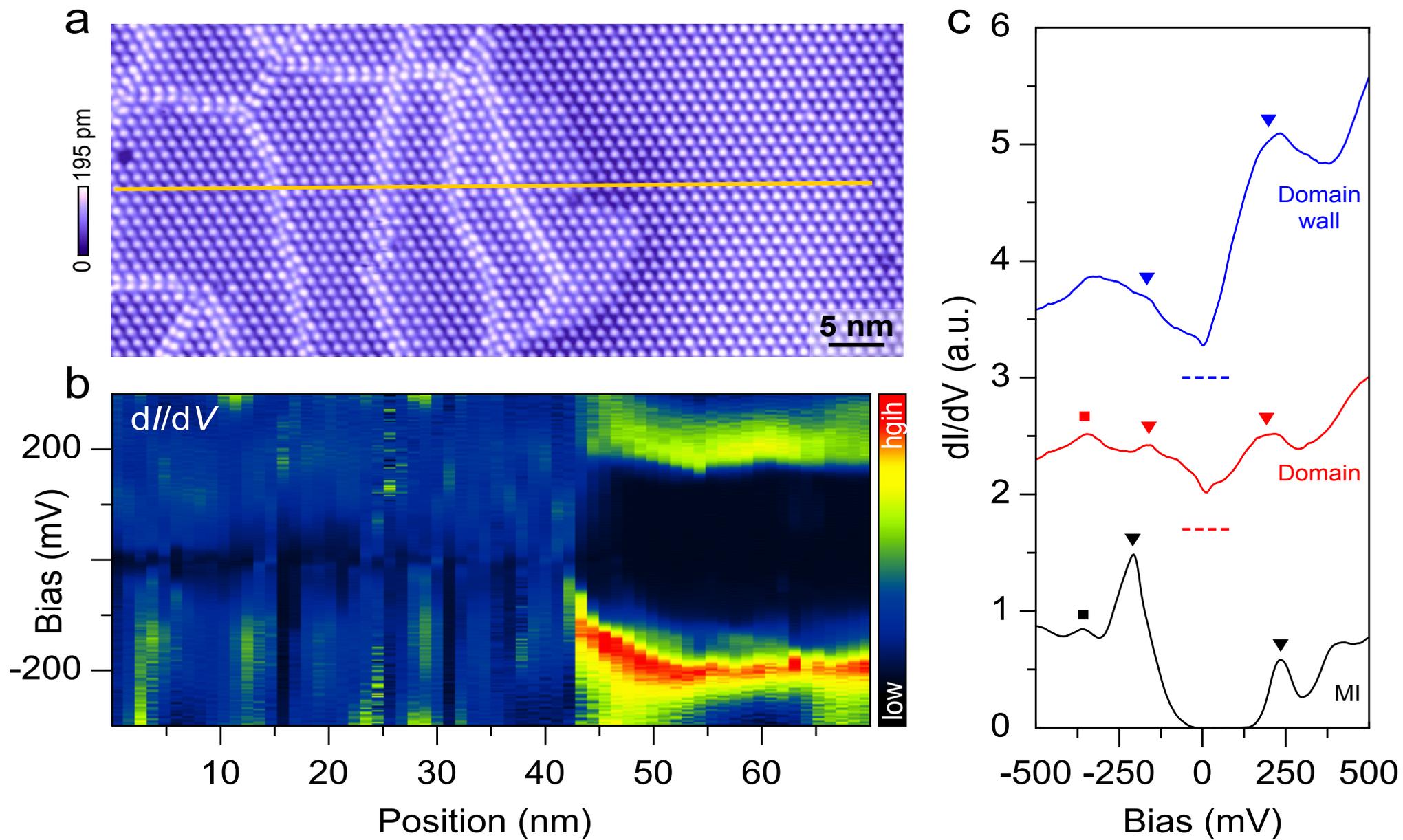

Liguo Ma *et al*. Figure 2

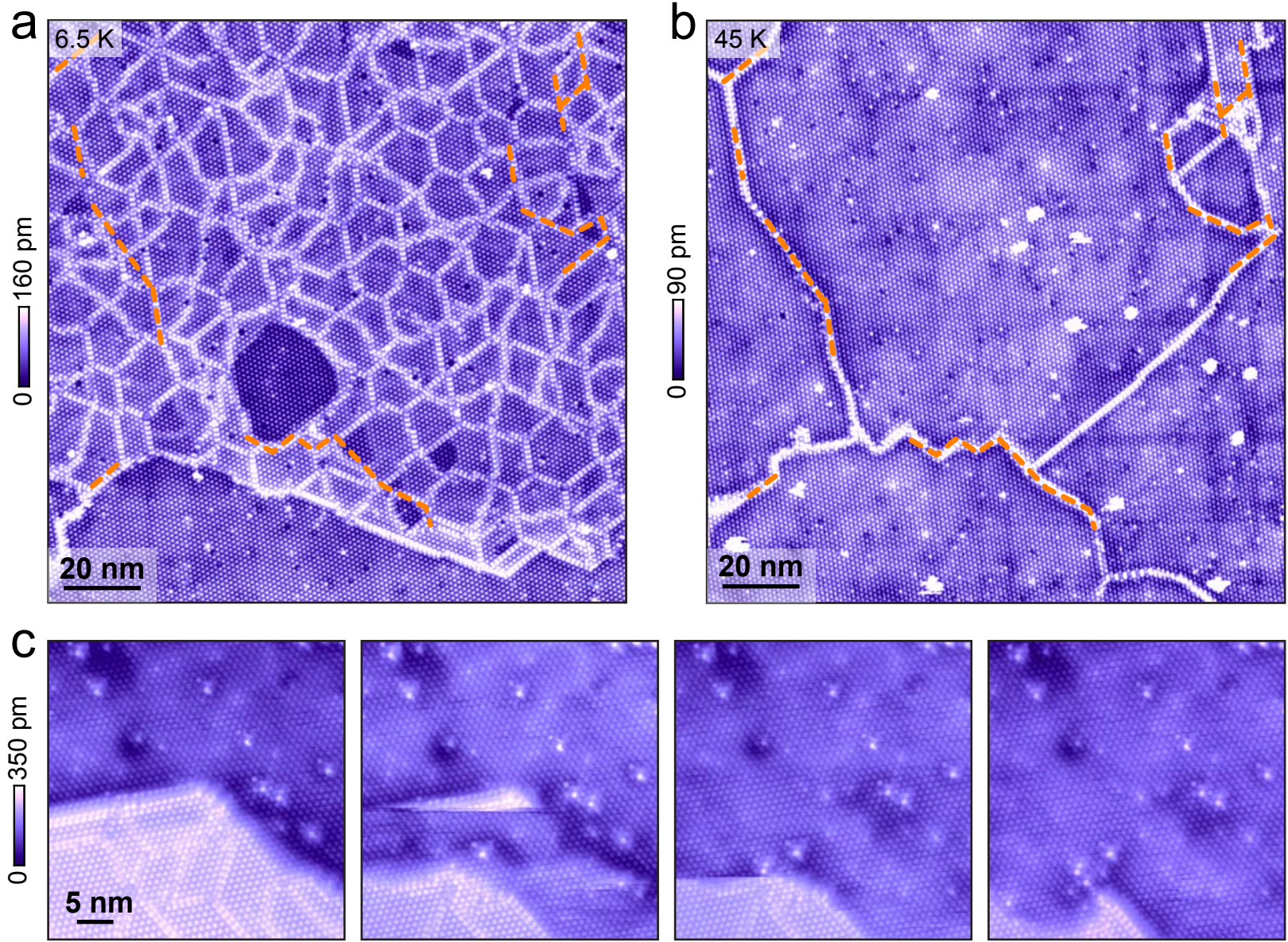

Liguo Ma *et al*. Figure 3

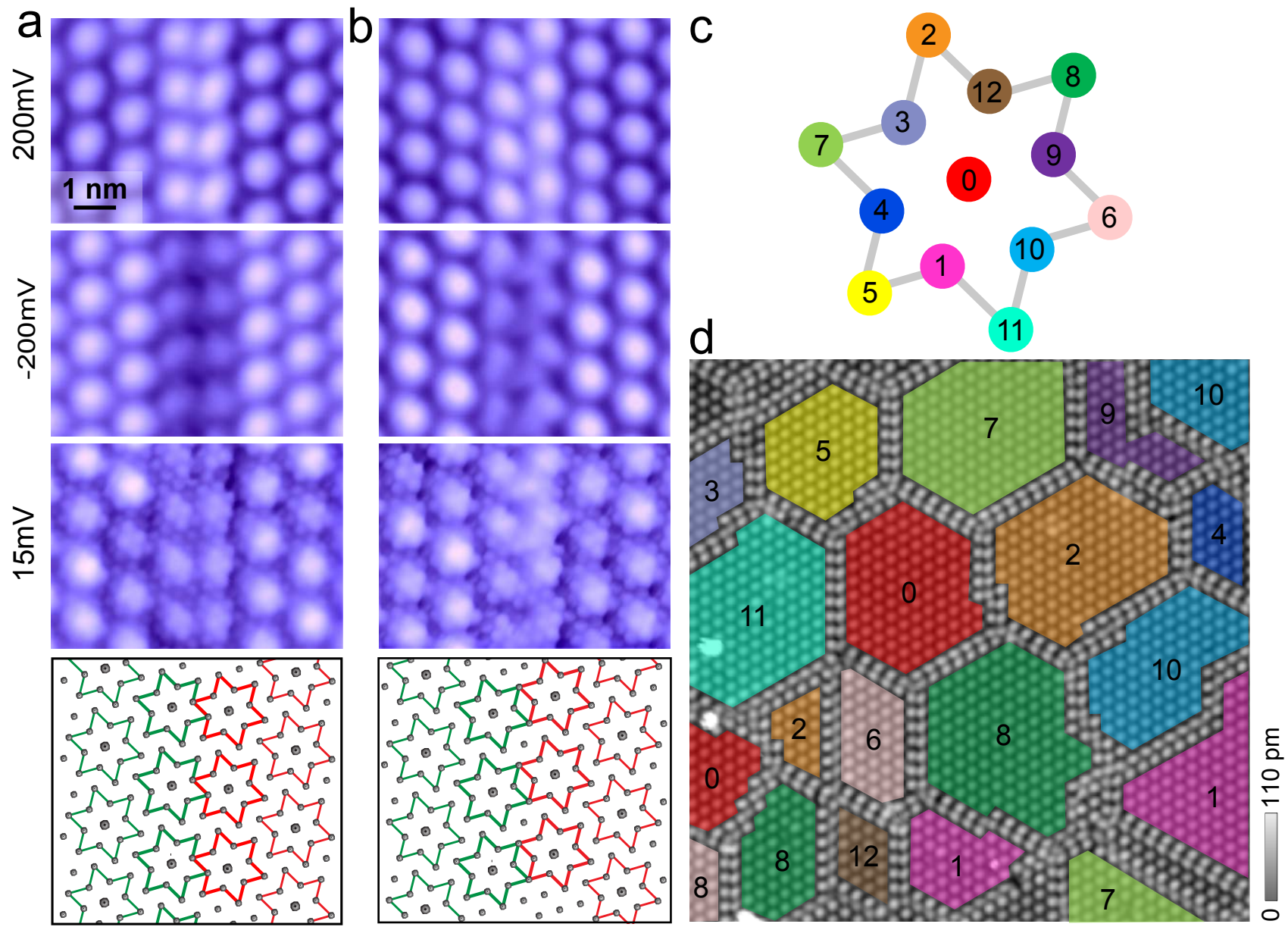

Liguo Ma *et al*. Figure 4

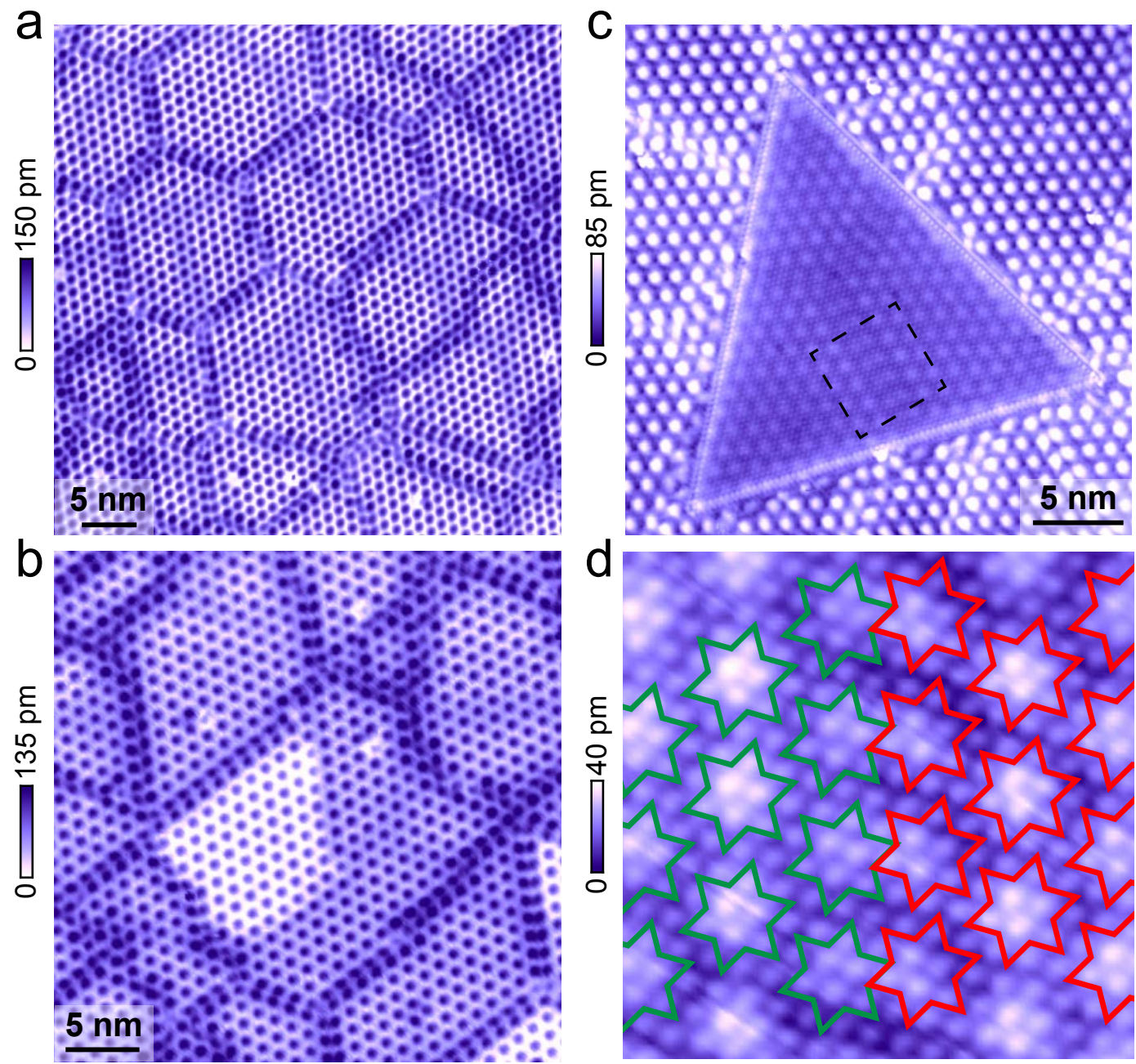

Liguo Ma *et al*. Figure 5

# Supplementary Information for

# A metallic mosaic phase and the origin of Mott insulating state in 1T-TaS$_2$

Liguo Ma, Cun Ye, Yijun Yu, Xiu Fang Lu, Xiaohai Niu, Sejoong Kim, Donglai Feng, David Tománek, Young-Woo Son, Xian Hui Chen and Yuanbo Zhang[*]

*Email: zhyb@fudan.edu.cn

## Content

# I. Angle-resolved photoemission spectroscopy (ARPES) characterisation of 1T-TaS$_2$ single-crystal

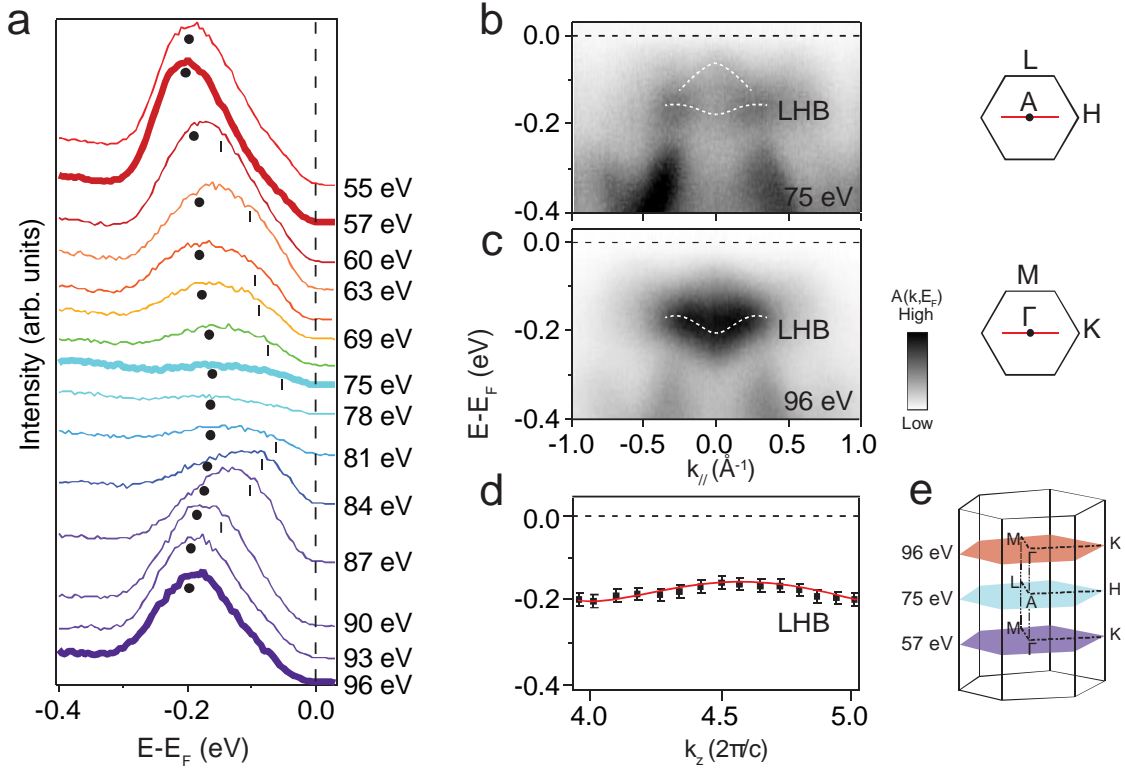

**Supplementary Figure 1 | ARPES characterisation of single-crystal 1T-TaS$_2$. a,** Normal emission ARPES spectra of 1T-TaS$_2$ taken in the CCDW phase at $T = 30$ K. The photon energy varies from 55 eV to 96 eV. The strong intensity pileup (marked by black circles) at around 200 meV below $E_F$ corresponds to the lower Hubbard band (LHB), and an additional dispersive peak (marked by ticks) appears inside of the Mott-Hubbard gap. Both features agree with those observed in ref. S1. The photon-energy-dependent spectra map out the dispersion of the energy bands in $k_z$ direction, with the purple, blue and red curves corresponding to colour-coded high-symmetry planes in $k$ space shown in **e**. We did not observe energy bands that cross the Fermi level in $k_z$ direction, in contrast to recent predictions based on *ab initio* calculations[S2]. **b** and **c,** Energy bands of 1T-TaS$_2$ along the A − H and Γ − K direction (indicated in the 2D Brillouin zone shown on the right), respectively. White dashed lines are guides to the eye showing the dispersion of the LHB along the two directions. From the dispersion we obtain the in-plane bandwidth of the LHB, $w_\parallel \approx 50$ meV. **d,** $k_z$ dispersion of the LHB extracted from **a**. The bandwidth of LHB along $k_z$ direction, $w_\perp \approx 40$ meV, is comparable to the in-plane bandwidth. **e,** Brillouin zone of 1T-TaS$_2$ showing high-symmetry planes and their corresponding photon energy. High-resolution ARPES measurements were performed at the SIS beamline of Swiss Light Source (SLS) equipped with a Scienta R4000 electron analyser. The overall energy resolution was better than 20 meV, and the angular resolution was 0.3 degrees. Samples were cleaved in ultra-high vacuum. ARPES measurements for each sample were carried out within 8 hours, with the sample aging effects carefully monitored.

## II. Mechanism of the tip-induced phase-switching

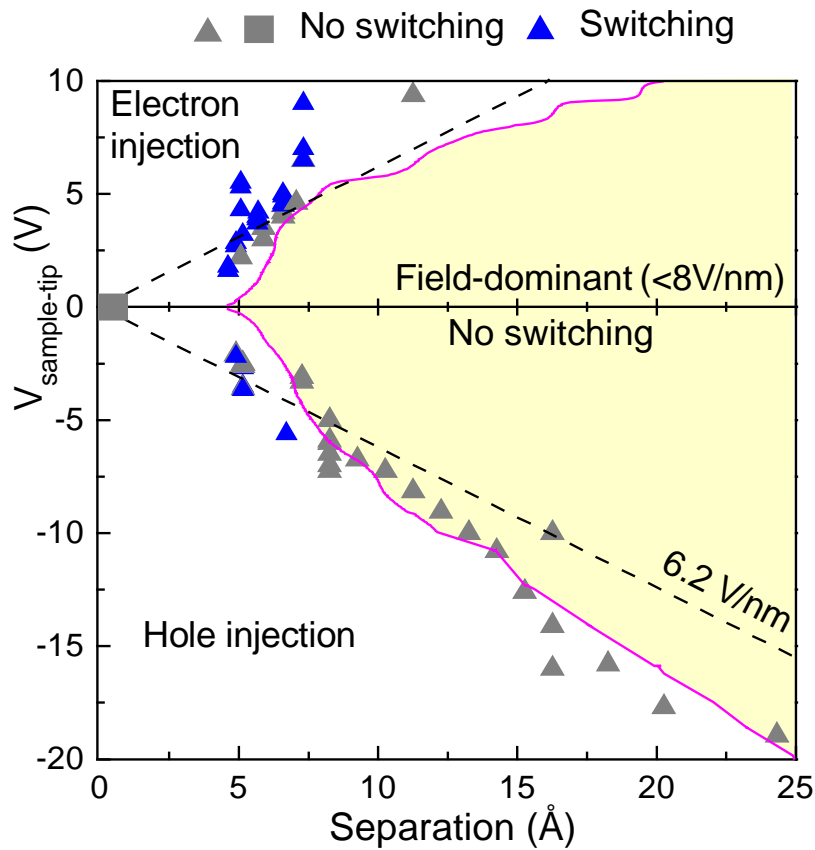

**Supplementary Figure 2 | Statistics of experimental attempts of inducing phase-switching with controlled tip-sample voltage and separation.** The magenta line corresponds to the tunnelling condition under which the tunnelling current between the sample and tip is fixed at 1nA. The enclosed parameter space (marked yellow) therefore corresponds to tunnelling with current less than 1nA. We found that continued scanning in yellow region did not produce any phase switching, even if the electric field is increased up to 8 V/nm. (The grey triangles denote individual pulses that did not produce phase switching either.) Meanwhile, pulses with similar electric field but smaller tip-sample separation (*i.e.* larger tunnelling/emission current; up to 20 nA in our experiment) can trigger the switching (blue triangles). We thus conclude that a pure electric field (without current) cannot switch the phase. We further studied the effect of a pure current injection (up to 100 uA) through the point contact between STM tip and sample without the presence of a large electric field (grey square), and didn't observe phase switching. This rules out ohmic joule heating as the driving force of the phase switching. Because both high electric field and current are required to induce the phase switching, we speculate that sudden local heating by hot electron/holes, followed by quenching, causes the switching (Ref. S3 and Fig. S10).

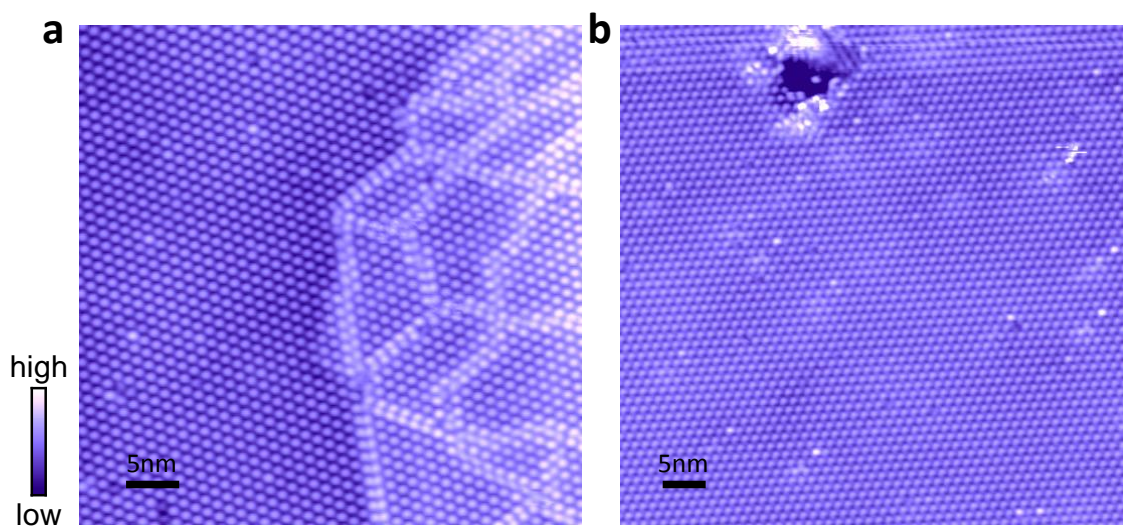

**Supplementary Figure 3 | MM phase induced by negative voltage pulse and the lack of phase switching from mechanical damage. a,** STM image of a MM patch induced by a −3.6 V pulse. **b,** STM image after mechanically crashing the STM tip onto the sample. MI to MM transition was not observed.

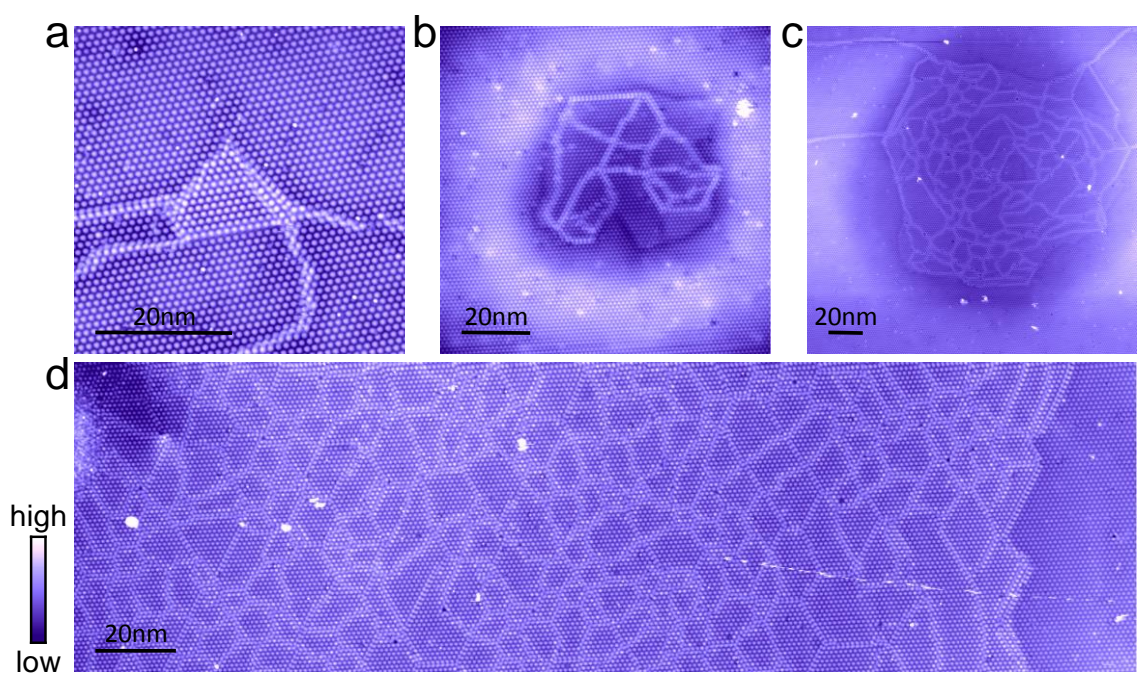

**Supplementary Figure 4 | Controlling the size of the MM patches by pulse voltage. a-d,** STM images of MM patches created by voltages pulses of 1.6, 2.7, 3.2 and 9.0 V, respectively. The diameters of the MM patches in **a**-**d** are about 15 nm, 50 nm, 130 nm, 500 nm (only part of the patch is shown here), respectively.

# III. Atomic structure of the domain walls in the MM phase

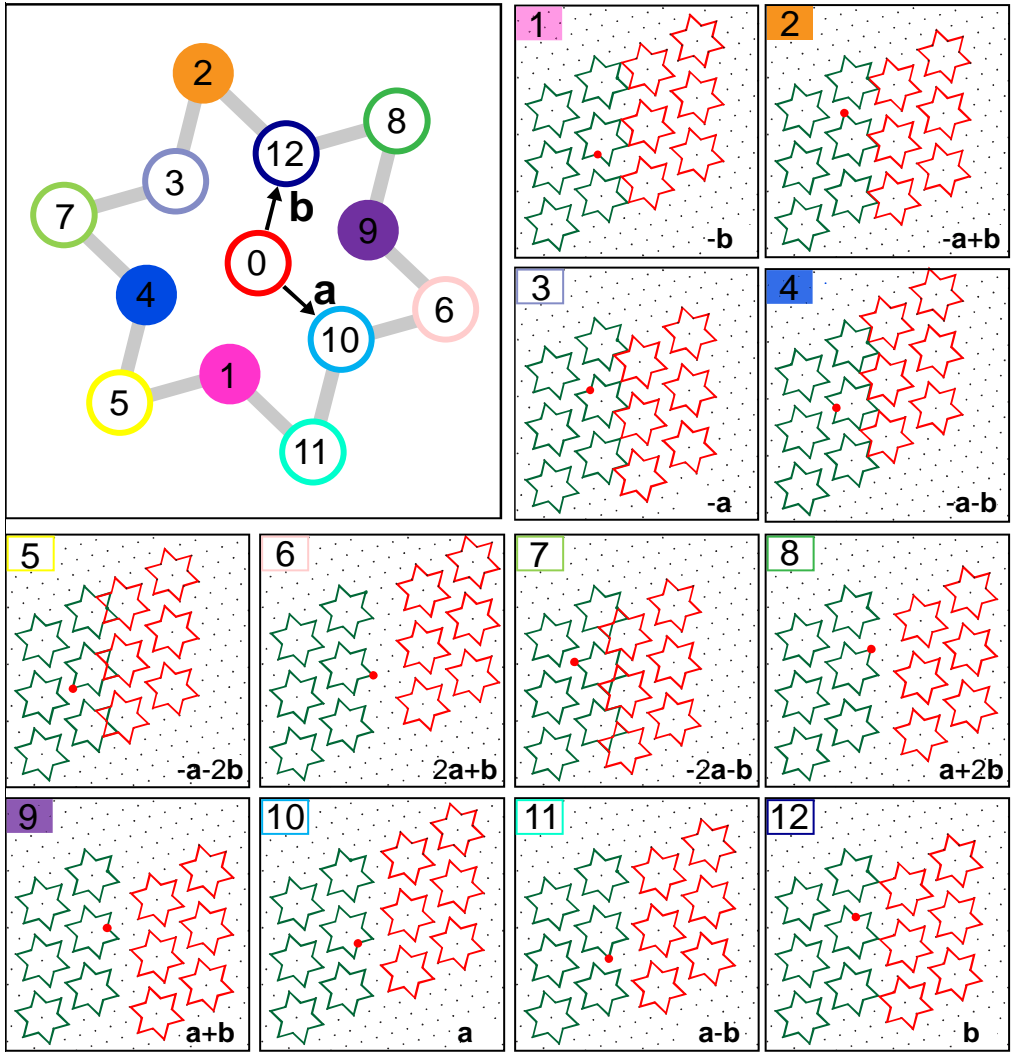

**Supplementary Figure 5 | Twelve possible boundary configurations between CCDW domains.** Upper-left panel: the thirteen Ta atoms in a David-star unit cell of the CCDW phase of 1T-TaS$_2$. $a$ and $b$ are unit vectors of the underlying atomic lattice. (unit vectors of the CCDW superlattice can then be written as $A = 4a + b$ and $B = 3b - a$.) Panel **1**-**12** show the twelve possible domain wall configurations. Assuming the underlying atomic lattice remain intact and there is no relative rotation of the CCDW order, the twelve domain wall configurations shown here exhaust all possibilities. Each possible configuration can be obtained by translating the David-star pattern (red) relative to that of the adjacent domain (green), so that the centre of the David-star, up to a CCDW superlattice vector, coincides with one of the twelve surrounding Ta atoms (marked by the red dot in panel 1-12; number of the atom shown on the upper-left corner). Here we adopt the convention used in ref. S4 to number the Ta atoms, which has the advantage that two consecutive translation can be represented by the difference of the two numbers. The corresponding shift of the David star centre, $ma + nb$ ($m$ and $n$ take integer values), is indicated in each panel. Each domain wall configuration, therefore, is characterized by the shift in phase of the CDW order parameter, $(\Delta\theta_1, \Delta\theta_2, \Delta\theta_3) = 2\pi(\frac{3m+n}{13}, \frac{-4m+3n}{13}, \frac{m-4n}{13})$. The numbers on solid colour indicate the domain wall configurations that are observed in the experiment (Fig. S6).

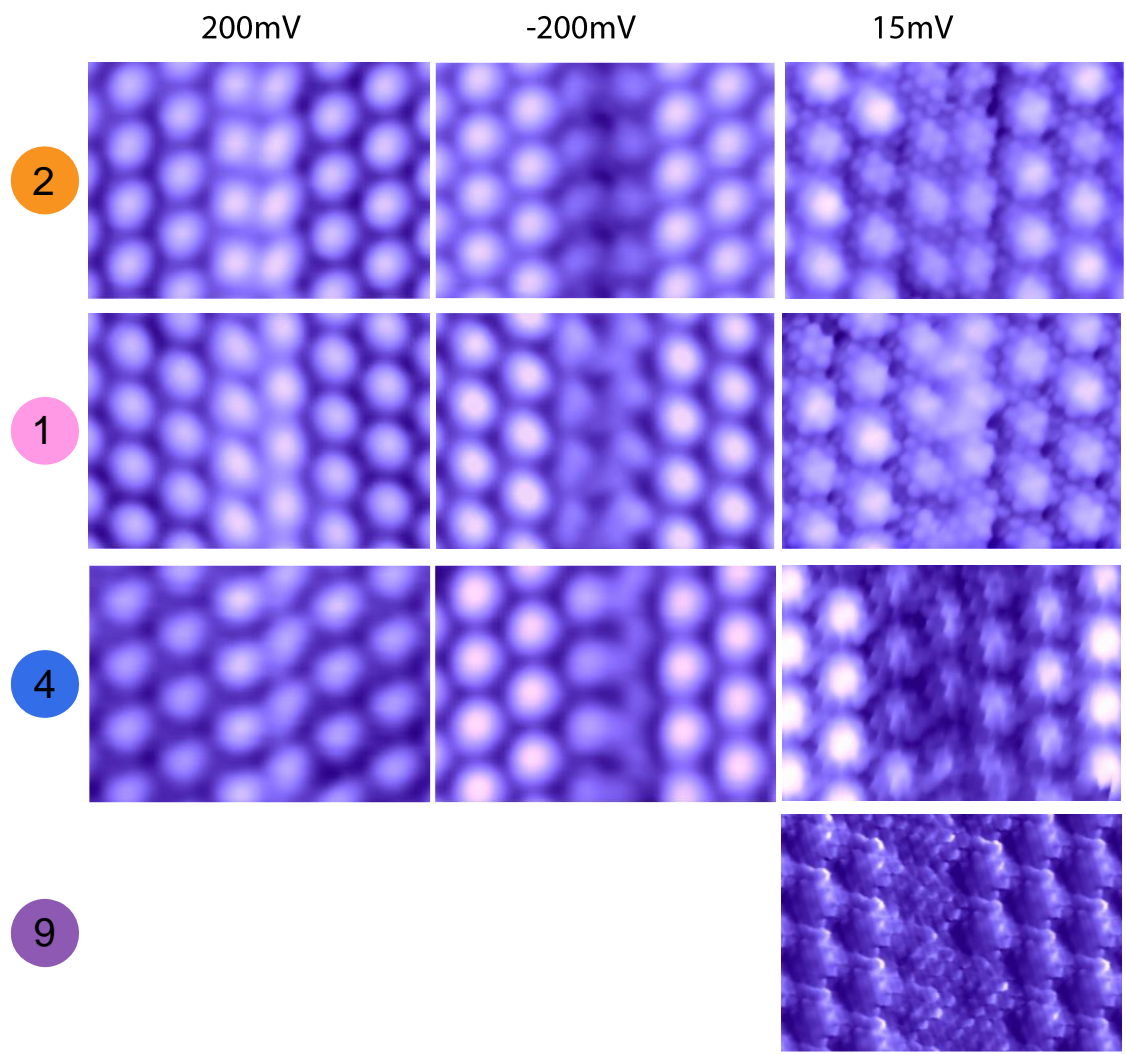

**Supplementary Figure 6 | Four types of domain walls observed in our experiments.** STM images of all four types of domain walls observed in our experiments under sample bias of 200 mV, -200 mV and 15 mV, respectively. All the experimentally observed domain walls are described by the atomic configurations (**2**, **1**, **4** and **9**, specifically) discussed in Fig. S5. The first two (**2** and **1**) are two most common types of domain walls seen in the experiment, and the last one (**9**) is rarely seen.

# IV. Conductive domain boundaries inside of the Mott insulating CCDW phase

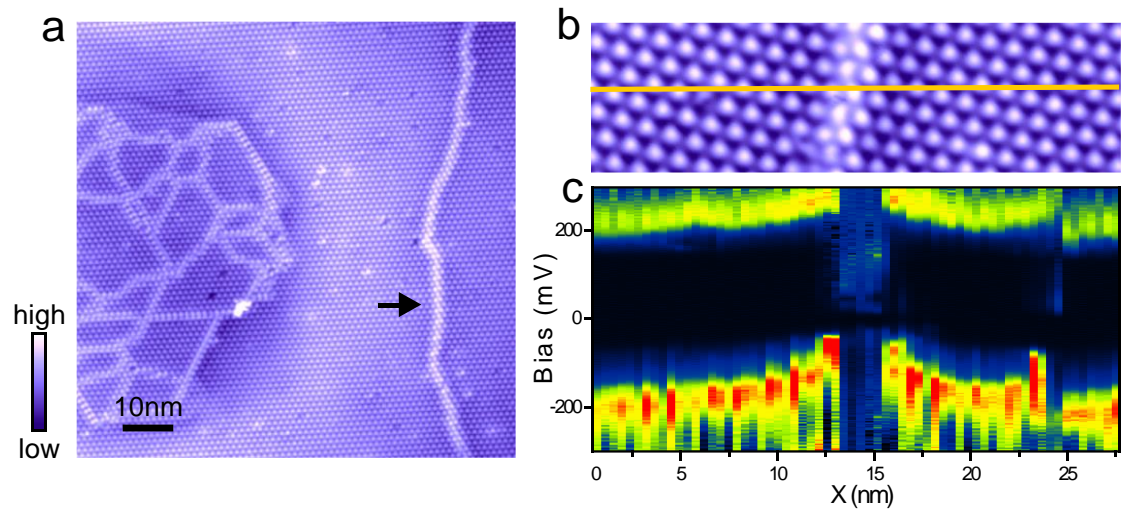

**Supplementary Figure 7 | A conductive domain boundary in insulating CCDW phase induced by voltage pulse. a**, STM image of a single domain boundary (indicated by the black arrow) and the MM phase, both of which are induced by a voltage pulse. Such domain boundaries inside of CCDW phase are occasionally observed after voltage pulses. The length of the boundary is typically on the order of ~ 100 nm. **b,** Zoomed-in STM image of the domain boundary in **a**. The boundary also corresponds to a phase shift in the phase of the CDW order parameter. **c,** $dI/dV$ spectrum along the orange line in **b**. The Mott-Hubbard gap disappears at the domain boundary, and the boundary exhibits a metallic behaviour.

## V. Bending of Mott-Hubbard bands at domain boundaries

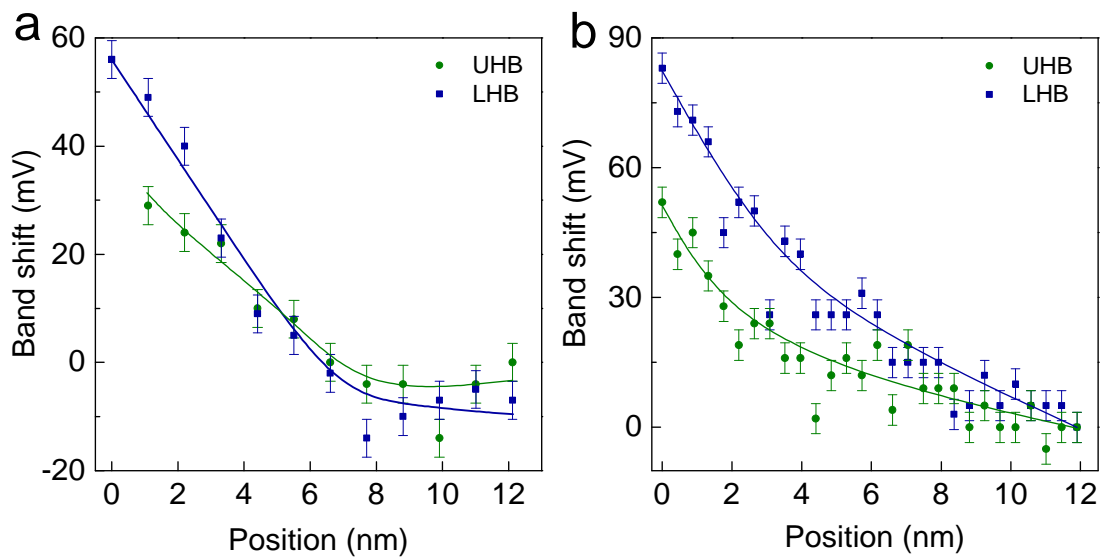

**Supplementary Figure 8 | Band bending at MM-MI interface and at conductive boundary in CCDW phase. a,** Bending of the upper and lower Hubbard bands near the interface between the MM phase and MI phase. The position is relative to the interface where the gap closes. Solid lines are guide for the eye. **b,** Bending of the upper and lower Hubbard bands near the conductive boundary in insulating CCDW phase discussed in Fig. S8. The band-bending behaviour is similar to that at the metal-semiconductor interface[S8], where the result is the combined effect of metal-induced gap states and bond polarization at the interface. However, further study is needed to understand the mechanism of the bending of the Hubbard bands.

## VI. Nearly commensurate CDW phase of 1T-TaS$_2$

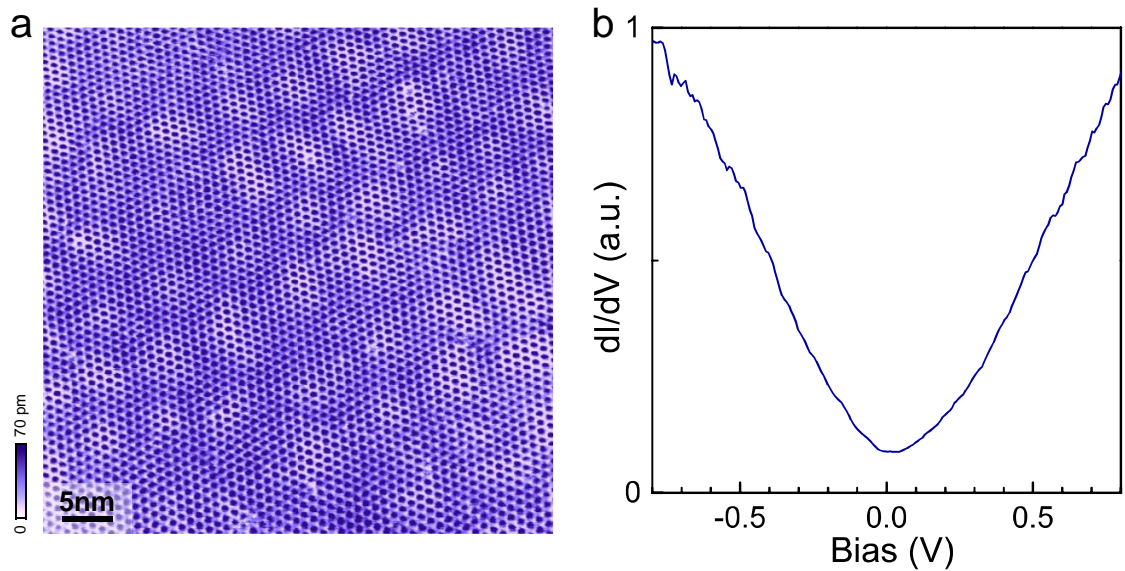

**Supplementary Figure 9 | Nearly commensurate CDW (NCCDW) phase of 1T-TaS$_2$ at room temperature. a,** STM image of NCCDW phase at 300K. CCDW domains are separated by incommensurate domain wall networks. The CDW phase order parameters are different between domains while the amplitude was smoothly modulated from domain centre to domain wall. **b,** $dI/dV$ spectrum acquired on surface of NCCDW phase at 300K. The induced MM phase is similar to NCCDW phase. They both are conductive and exhibiting a domain-like pattern. The average domain size of MM phase is 80 nm$^2$, about the same with that of NCCDW phase[S9] near the NC to C transition point (200K). The difference is NCCDW phase has distorted Kagome lattice domains[S10] (or so called corner sharing hexagonal domains) with broadened domain walls while complex domain distribution with sharp domain walls in MM phase.

## VII. Super-cooled NCCDW phase of 1T-TaS$_2$

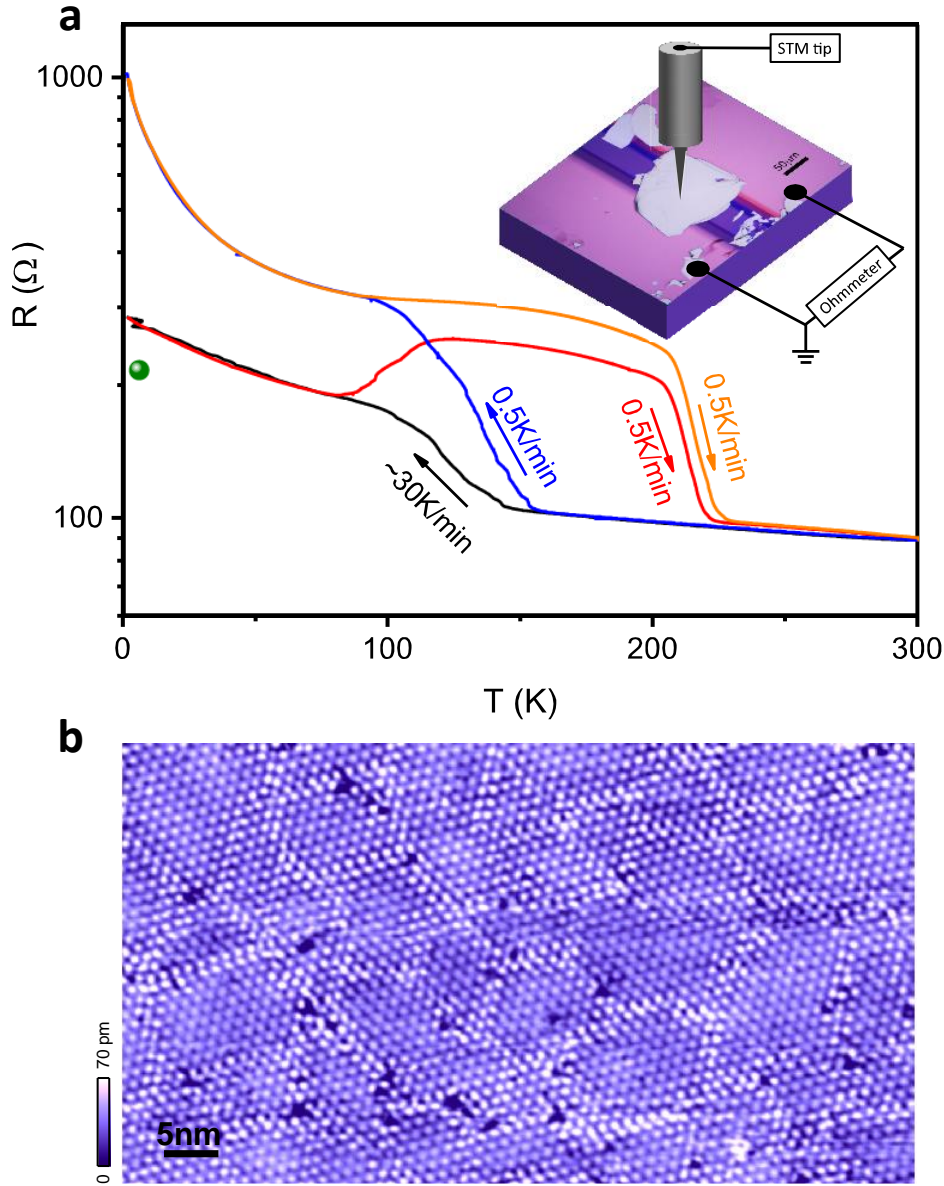

**Supplementary Figure 10 | Transport and STM characterisation of the super-cooled NCCDW phase. a,** Resistance as a function of temperature of a 1T-TaS$_2$ thin flake under various cool-down and warm-up rate. The CCDW state is developed at low temperature under a low cool-down rate (blue curve), whereas under a high cool-down rate a super-cooled NCCDW state is observed [S11] (black curve). **b,** STM image of the super-cooled NCCDW state obtained on the same device as shown in the inset of **a** (imaging condition: $V_t = 1$ V and $I_t = 60$ pA). The super-cooled state was realized by quenching the device from room temperature to liquid helium temperature (with an effective cooling rate of ~ 70 K/min). The low-temperature sample resistance is shown in **a** (green dot), indicating that the sample is indeed in the super-cooled NCCDW state. The topography of the super-cooled NCCDW phase shows randomly distributed domains, which are similar to those in the pulse-induced MM phase.

# VIII. Calibration of STM tip before and after STS measurements on TaS$_2$

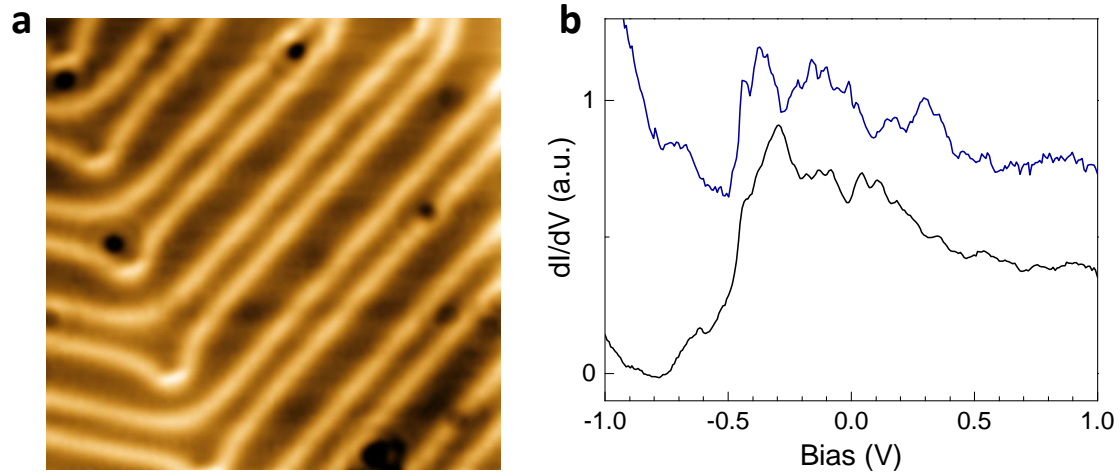

**Supplementary Figure 11 | Spectral calibration of the STM tip on Au(111) surface before and after measurements on 1T-TaS$_2$.** **a,** Constant-current STM image of Au (111) surface ($V_s = 0.5$ V, $I_t$=10 pA) showing herringbone reconstruction. **b,** Typical *dI/dV* spectrum measured on Au (111) surface before (black) and after (blue) the STM tip was used to measure 1T-TaS$_2$ samples. All the STM tips used in this study were calibrated on clean Au(111) surface prior to measurement on 1T-TaS$_2$ to ensure that the tips show Au(111) surface state at $V_s = -0.5$ V, similar to that in ref. S12. The tips were checked again on Au(111) surface after the measurement on 1T-TaS$_2$ to confirm that the surface state can be reproduced. This calibration procedure ensures that the tip does not have anomalous spectroscopic features during the measurement on 1T-TaS$_2$.

## IX. First-principle calculation of energies associated with different stacking order

Fully relaxed David-star geometry for a single and bilayer 1T-TaS$_2$ has been obtained through density functional theory (DFT) calculation by using Quantum Espresso Code[S13]. We adopted the PBE generalized gradient approximation[S14] for exchange-correlation functional, the norm-conserving pseudo-potential[S15], an energy cutoff of 55 Ry and 6x6 *k*-points for the supercell. In the relaxed bilayer 1T-TaS$_2$, the center of David-star in the first (upper) layer is found to shift by **a** relative to the center in the second (lower) layer. For trilayer 1T-TaS$_2$, the binding energy is calculated for two cases: 1) the center of the third (bottom) layer David-star center shifts by 2**a** with respect to the center in the first (topmost) layer, *i.e.* ABC stacking and 2) the center in the third layer is same to one in the topmost layer, *i.e.* ABA stacking. We computed the energy difference as the David-star in one layer is displaced relative to the other layers, and found the difference is about 6.4 meV.